\documentclass[11pt,twocolumn,twoside]{IEEEtran}
\usepackage{amsmath}
\usepackage[margin=0.75in,headheight=0.45in]{geometry}
\usepackage[pdftex]{epsfig}
\usepackage{amsfonts}
\usepackage{amssymb}
\usepackage{fancyhdr}
\usepackage{xspace}
\usepackage[caption=false]{subfig}  
\usepackage{multicol}
\usepackage{hyperref}
\include{graphicsx}

\newcommand{\eM}     {\mbox{$\epsilon$-machine}\xspace}

\newcommand{\MeasSymbol}   { {X} }
\newcommand{\meassymbol}   { {x} }

\newcommand{\CausalState}       { \mathcal{S} }

\newcommand{\CausalEquivalence} { {\sim}_{\epsilon} }

\newcommand{\forward}{+}
\newcommand{\reverse}{-}
\newcommand{\forwardreverse}{\pm} 

\newcommand{\FutureCausalState} { {\CausalState}^{\forward} }

\newcommand{\PastCausalState}   { {\CausalState}^{\reverse} }

\newcommand{\lastindex}[2]{
  \edef\tempa{0}
  \edef\tempb{#2}
  \ifx\tempa\tempb
    \edef\tempc{#1}
  \else
    \edef\tempa{0}
    \edef\tempb{#1}
    \ifx\tempa\tempb
      \edef\tempc{#2}
    \else
      \edef\tempc{#1+#2}
    \fi
  \fi
  \tempc
}

\newcommand{\CSjoint}[1][,]{
   \edef\tempa{:}
   \edef\tempb{#1}
   \ifx\tempa\tempb
      \ensuremath{\FutureCausalState\!#1\PastCausalState}
   \else
      \ensuremath{\FutureCausalState#1\PastCausalState}
   \fi
}
\newcommand{\CSjointKL}[3][,]{
   \edef\tempa{:}
   \edef\tempb{#1}
   \ifx\tempa\tempb
      \ensuremath{\FutureCausalState_{#2}\!#1\PastCausalState_{#3}}
   \else
      \ensuremath{\FutureCausalState_{#2}#1\PastCausalState_{#3}}
   \fi
}

\newif\ifpm
\edef\tempa{\forwardreverse}
\edef\tempb{\pm}
\ifx\tempa\tempb
   \pmfalse
\else
   \pmtrue
\fi

\ifcsname mathclap\endcsname
  \relax
\else
  \def\clap#1{\hbox to 0pt{\hss#1\hss}}

\fi

\newcommand{\pastt} {\overleftarrow{\meassymbol}}
\newcommand{\Pastt} {\overleftarrow{\MeasSymbol}}

\newcommand{\Futuree} {\overrightarrow{\MeasSymbol}}

\newcommand{\Localstate} {Local causal state\xspace}
\newcommand{\Localstates} {Local causal states\xspace}
\newcommand{\localstate} {local causal state\xspace}
\newcommand{\localstates} {local causal states\xspace}
\newcommand{\stpoint} {(\vec{r}, t)}

\begin{document}
\title{\vspace{0.2in}\sc A Physics-Based Approach to Unsupervised Discovery of Coherent Structures in Spatiotemporal Systems}
\author{Adam Rupe$^{1,2}$\thanks{Corresponding author: A Rupe, atrupe@ucdavis.edu $^1$Complexity Sciences Center, Department of Physics, University of California Davis $^2$NERSC, Lawrence Berkeley National Laboratory}, James P. Crutchfield$^{1}$, Karthik Kashinath$^{2}$, Mr Prabhat$^{2}$}

\maketitle
\thispagestyle{fancy}
\begin{abstract}
Given that observational and numerical climate data are being produced at ever more prodigious rates, increasingly sophisticated and automated analysis techniques have become essential. Deep learning is quickly becoming a standard approach for such analyses and, while great progress is being made, major challenges remain. Unlike commercial applications in which deep learning has led to surprising successes, scientific data is highly complex and typically unlabeled. Moreover, interpretability and detecting new mechanisms are key to scientific discovery. To enhance discovery we present a complementary physics-based, data-driven approach that exploits the causal nature of spatiotemporal data sets generated by local dynamics (e.g. hydrodynamic flows). We illustrate how novel patterns and coherent structures can be discovered in cellular automata and outline the path from them to climate data.
\end{abstract}

\section{Motivation}
Incredibly complex and sophisticated models are currently employed to simulate the global climate system to facilitate our understanding of climate as well as increase our predictive power, most notably in regards to the effects of increased carbon levels. Our ability to simulate however has rapidly outpaced our ability to analyze the resulting data. Often the climate community resorts to rather simplistic data analyses, such as linear decomposition methods like EOF analyses \cite{NCARa, Ghil02a} or detecting (linear) trends in climate data time series \cite{NCARb}. Nonlinear and more sophisticated techniques are rarely brought to bear. Here we focus on one particular aspect of nonlinear dynamical systems analysis, the detection and discovery of coherent structures, such as cyclones and atmospheric rivers in climate data. 

Coherent structures were introduced in the study of fluid dynamics and were initially defined as regions characterized by high levels of coherent vorticity, \textit{i.e.} regions where instantaneously space and phase correlated vorticity are high. 
The contours of coherent vorticity constitute an identifier to the structure's boundaries. 
However, pinning down this concept of coherent structures with rigorous and principled definitions or heuristics which produce consistent results across a wide class of physical systems is a challenging and open problem \cite{Hadj17a}.
Climate practitioners are left with more ad hoc approaches \cite{Vita97a,Wals97a,Prab15a} which can make it difficult to draw meaningful conclusions from analysis \cite{Fara16a}.

Deep learning attempts to sidestep this issue by learning how to identify coherent structures from labeled data \cite{Liu16a}. 
However, we currently can not peer into the box to find out exactly what the defining characteristics a deep net uses to identify structures. Current state of the art achieves semi-supervised bounding box identification \cite{Raca16a}. The ultimate goal would be unsupervised segmentation; that is, a pixel-level identification without reliance on labeled training data. It is not yet clear how to achieve this. 

Like deep learning, our theory \cite{Rupe17a} approaches coherent structures from a rather different (and more general) perspective than the original context of Lagrangian coherence principles in fluid flows. 

\section{Method}
Starting from basic physics principles, coherent structures can most generally be seen as \emph{localized broken symmetries}. Two questions naturally arise; what are the symmetries which are broken and how can we identify such symmetry in a diverse range of spatiotemporal systems? Coherent structures can be found in a variety of systems with different physical properties. Convection cells in hydrodynamic systems and spiral waves in reaction-diffusion systems, for example. It is clear that the common thread is the underlying nonlinear dynamics of these systems \cite{Cros09a,Hoyl06a,Golu03a}. 

A framework known as \emph{computational mechanics} \cite{Crut88a,Crut12a} has been developed to study pattern and structure in this dynamical context. The canonical object of computational mechanics is the \textit{\eM} \cite{Shal98a}, a type of stochastic finite-state machine known as a hidden Markov model, which consists of a set of \textit{causal states} and transitions between them. The causal states are constructed from the \textit{causal equivalence relation}.
\begin{align*}
\pastt_i \sim_\epsilon \pastt_j  \iff \Pr ( \Futuree | \Pastt = \pastt_i ) = \Pr ( \Futuree | \Pastt = \pastt_j).
\end{align*}
In words, two pasts $\pastt_i$ and $\pastt_j$ are \textit{causally equivalent} if and only if they make the same prediction for the future $\Futuree$; that is, they have the same conditional distribution over the future. The causal states are the unique \textit{minimal sufficient statistic} of the past to predict the future.

For our application to coherent structures we use a straightforward spatiotemporal generalization known as the \emph{local causal states} \cite{Shal03a}. For systems which evolve under some local dynamic and information propagates through the system at a finite speed, it is quite natural to use \textit{lightcones} as local notions of pasts and futures. Formally, the past lightcone of a spacetime point $\meassymbol(\vec{r}, t)$ is the set of all points at previous times that could possibly influence it. That is,
\begin{align*}
\ell^-(\vec{r}, t) \equiv \big \{\meassymbol(\vec{r}', t') \;|\; t' \leq t \; \mathrm{and} \; ||\vec{r}' - \vec{r}|| \leq c(t' - t) \big \}
\end{align*}
where $c$ is the finite speed of information propagation in the system. Similarly, the future lightcone is given as all the points at subsequent times that could possibly be influenced by $\meassymbol(\vec{x}, t)$.
\begin{align*}
\ell^+(\vec{r}, t) \equiv \big \{ \meassymbol(\vec{r'}, t') \;|\; t' > t \; \mathrm{and} \; ||\vec{r}' - \vec{r}|| < c ( t - t') \big \}
\end{align*}

The choice of lightcone representations for both local pasts and futures is ultimately a weak-causality argument; influence and information propagate locally through a spacetime site from its past lightcone to its future lightcone.

The generalization of the causal equivalence relation is straightforward. Two past lightcones are causally equivalent if they have the same conditional distribution over future lightcones. 
\begin{align*}
\ell^-_i \; \CausalEquivalence \; \ell^-_j \iff \Pr \big(\mathrm{L}^+ | L^- = \ell^-_i \big) = \Pr \big(\mathrm{L}^+ | L^- = \ell^-_j \big)
\end{align*}

This \textit{local causal equivalence relation} over lightcones is designed around an intuitive notion of \textit{optimal local prediction} \cite{Shal03a}. At some site $\meassymbol\stpoint$ in spacetime, given knowledge of all past spacetime points which could possibly affect $\meassymbol\stpoint$, i.e. its past lightcone $\ell^-\stpoint$, what might happen at all subsequent spacetime points which could be affected by $\meassymbol\stpoint$, i.e. its future lightcone $\ell^+\stpoint$? \Localstates are minimal sufficient statistics for optimal local prediction. Moreover, the particular local prediction done here uses lightcone shapes, which are associated with local causality in the system. Thus it is not direct causal relationships (e.g. learning equations of motion from data) that the \localstates are discovering. Rather, they are exploiting a kind of causality in the system  (i.e. that the future follows the past and that information propagates at a finite speed) in order to discover spacetime structure.

Once \localstates have been inferred from data, each site in a representative spacetime field can be assigned its \localstate label in a process known as \emph{causal filtering} \cite{Rupe17a}. This is how we achieve unsupervised image segmentation. Though it must be clearly stated that this is a \emph{spacetime segmentation}, and not a general image segmentation algorithm, exactly because it works \emph{only} in systems for which lightcones are well-defined.

Using the \localstates we can, in a general and principled manner, discover dynamical spatiotemporal symmetries in a system from data. These symmetry regions are known as \emph{domains} and are defined as regions where the associated \localstate field, after causal filtering, has spacetime symmetry tilings. A \emph{coherent structure} is then defined as a set of spatially localized, temporally persistent (in the Lagrangian sense) non-domain \localstates. 

From prior work by Hanson and Crutchfield \cite{Hans90a,Crut91d,Hans95a}, the domains of 1-D cellular automata are well understood as dynamically invariant sets of homogeneous spatial configurations. There is strong empirical evidence \cite{Rupe17a} that the domains of cellular automata discovered by the \localstates are \emph{exactly} the domains as described by Hanson and Crutchfield. Therefore the \localstates are discovering spatiotemporally symmetries which are externally well-defined. In turn there is a strong agreement between the description of coherent structures in cellular automata discovered by \localstates and the coherent structures as described by Hanson and Crutchfield. 

\begin{figure*}[h!]
  \begin{multicols}{2}
    \centering
	\subfloat[Raw spacetime field of ECA 54]{\epsfxsize=0.95\hsize \epsfbox{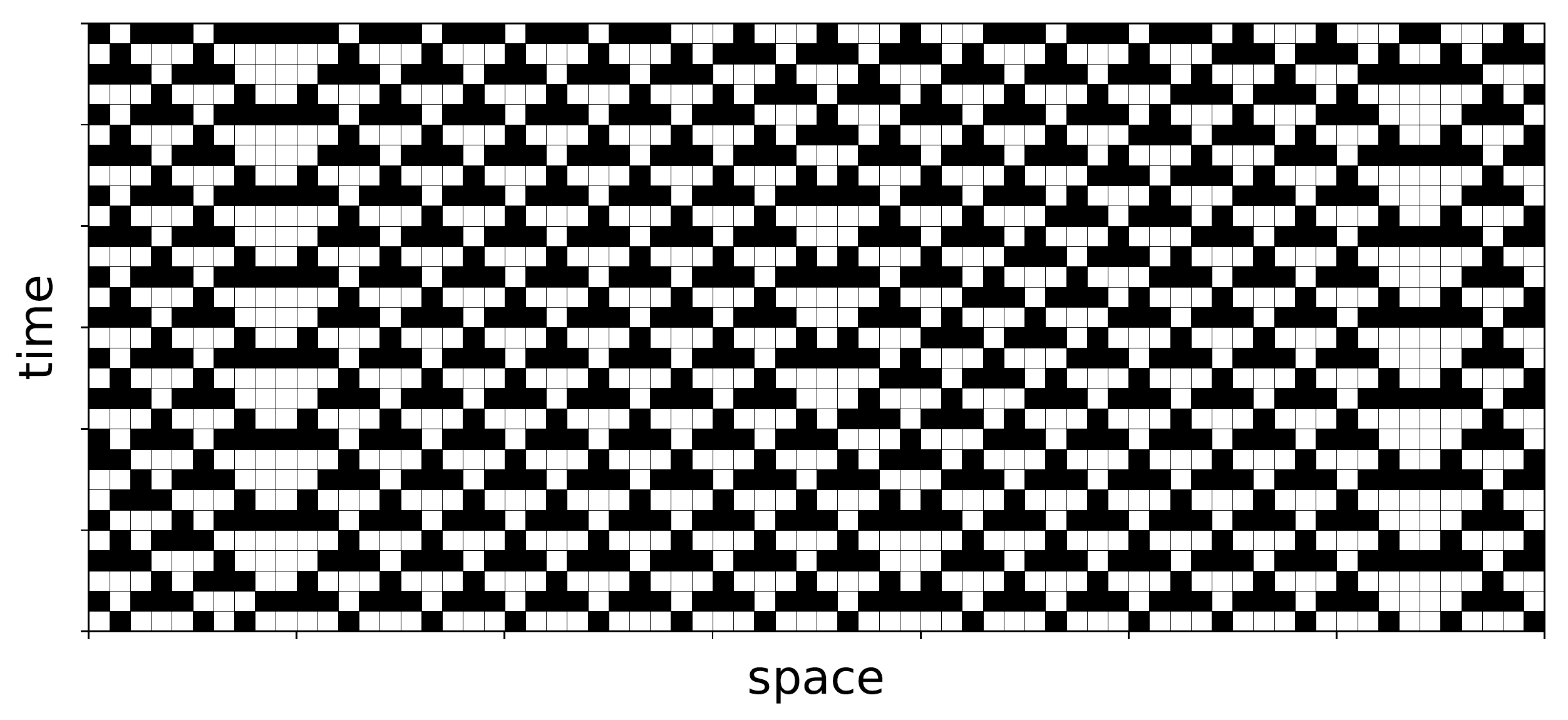}}\\
	\vspace{-3mm}
	\subfloat[Local statistical complexity filter of (a)]{\epsfxsize=0.95\hsize \epsfbox{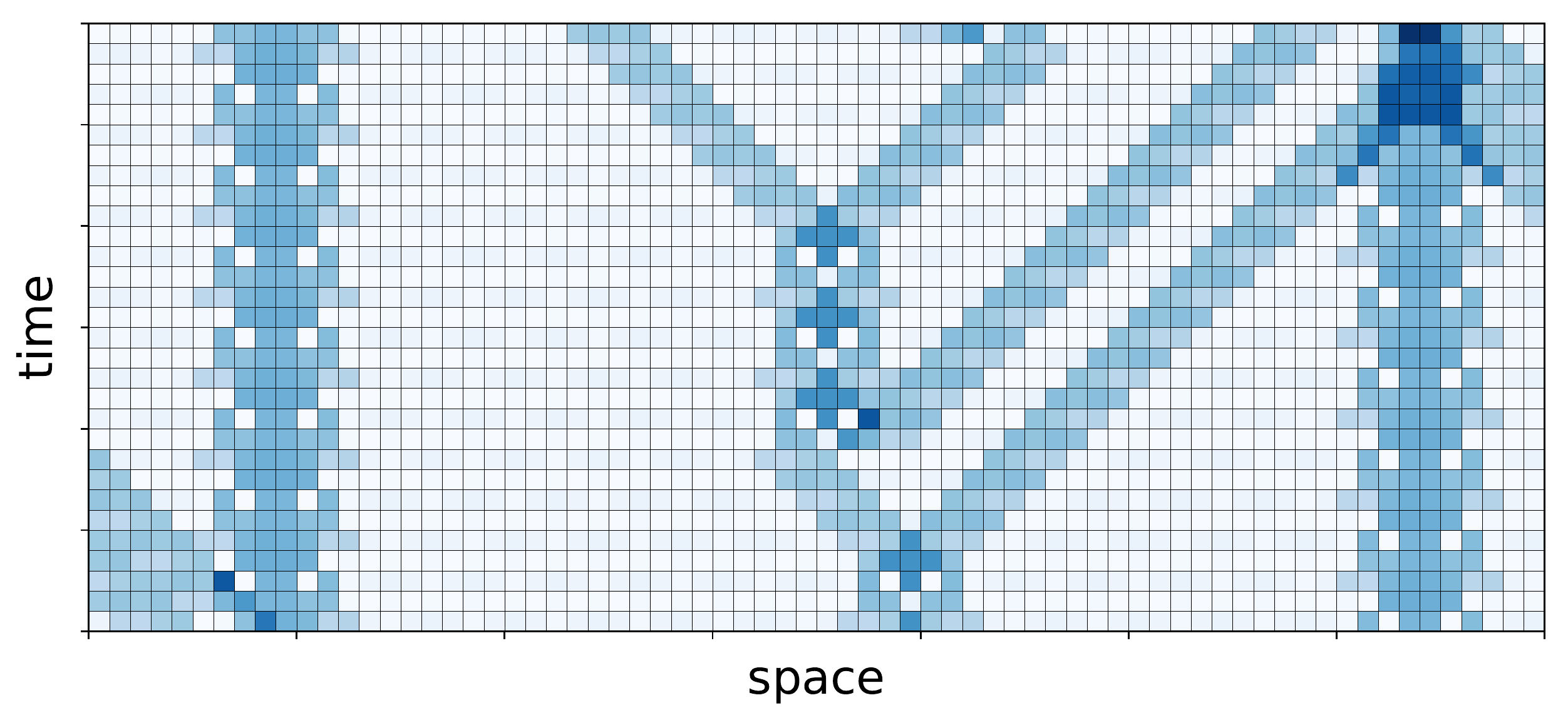}} \\
	\vspace{-3mm}
	\subfloat[\Localstate coherent structure filter of (a)]{\epsfxsize=0.95\hsize \epsfbox{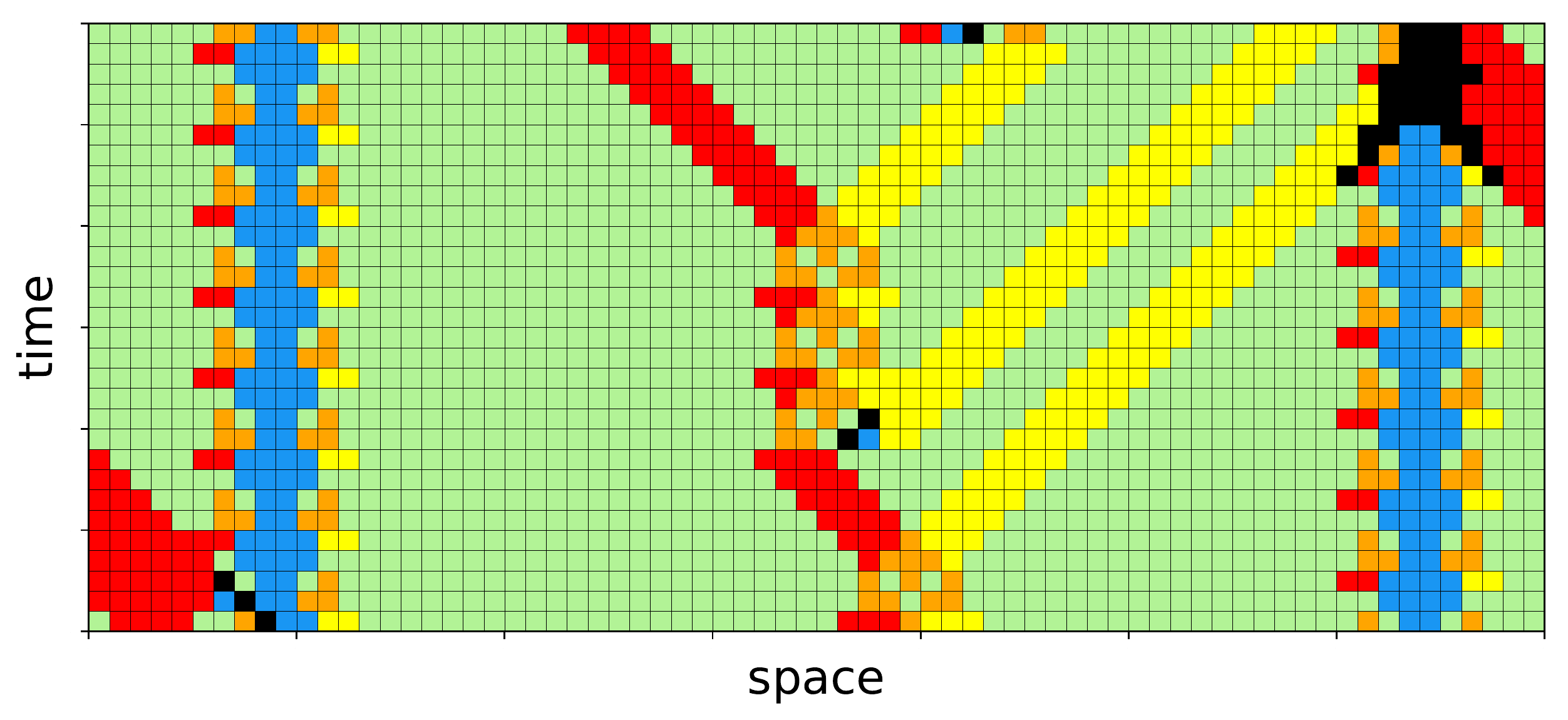}}\\
	\vspace{-3mm}

	\subfloat[Vortex street in atmosphere]{\epsfxsize=0.95\hsize \epsfbox{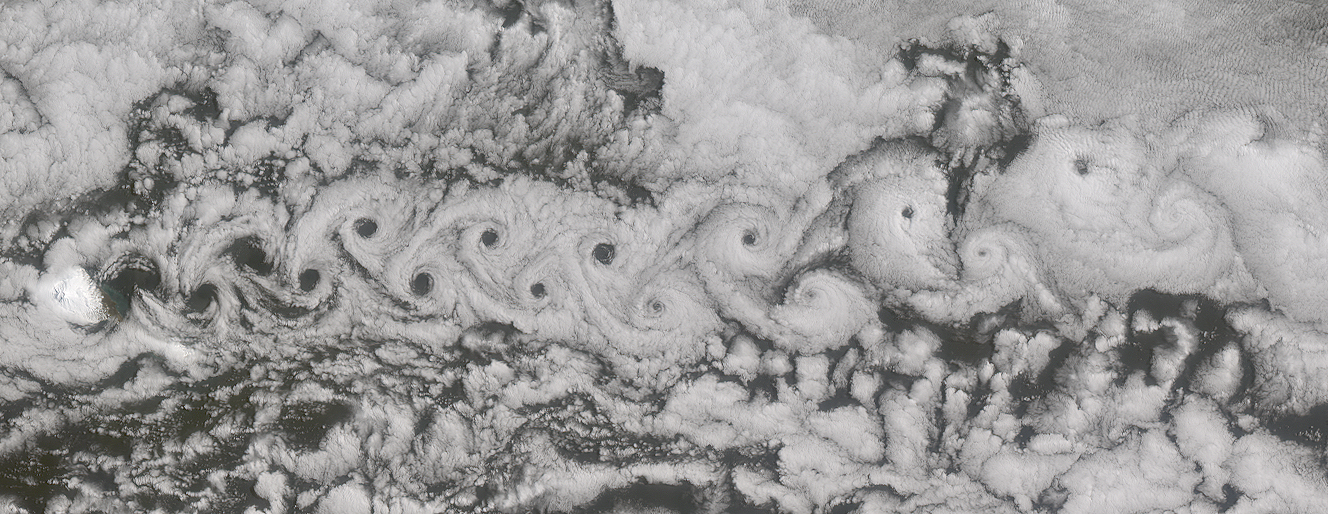}}\\
	\vspace{8mm}
	\subfloat[Vortex street complexity field]{\epsfxsize=0.95\hsize \epsfbox{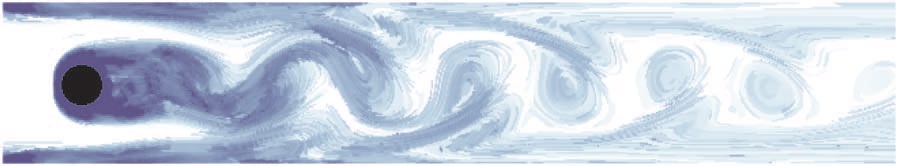}}\\
	\vspace{14mm}
	\subfloat[Colored vortices on Cori]{\epsfxsize=0.95\hsize \epsfbox{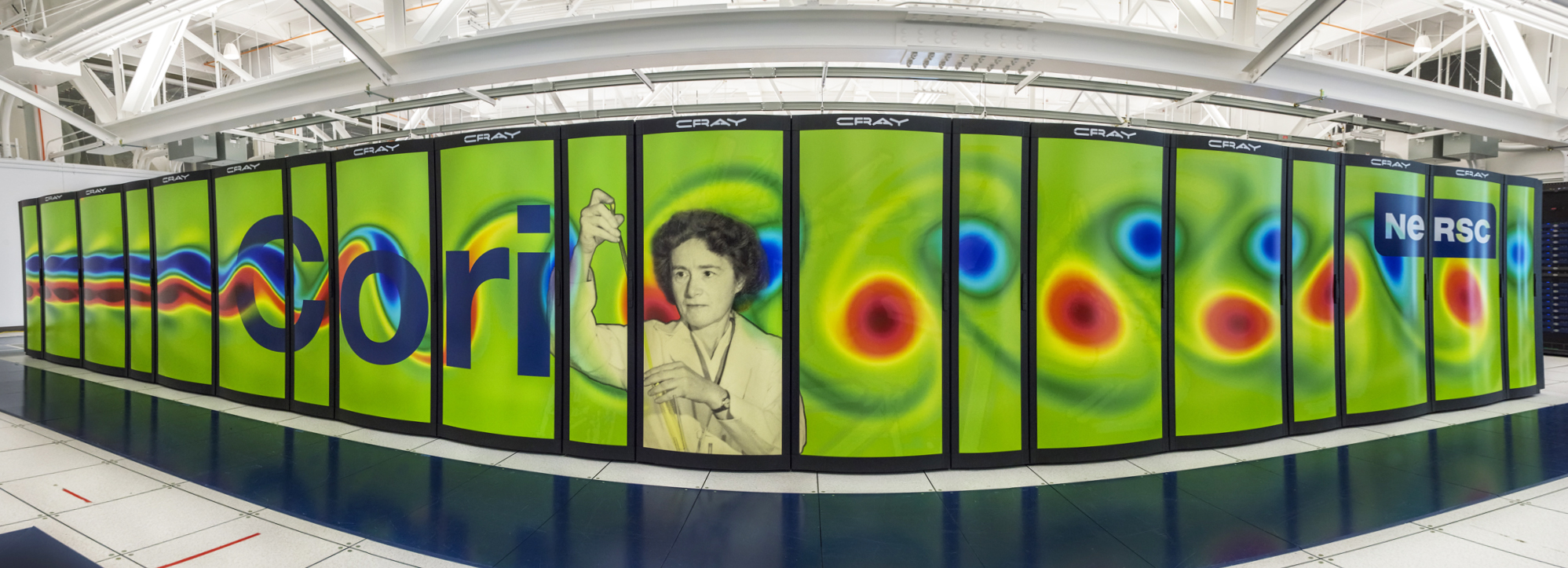}} 
  \end{multicols}
    \caption{Visualization of results on 1D cellular automata (fully-discrete spatiotemporal models) and projected analogous results for fluid systems. CA results for elementary cellular automaton rule 54 are given in (a)-(c). The raw spacetime field is shown in (a) and a corresponding local statistical complexity field in (b). From the local statistical complexity filter, which is a qualitative information-theoretic ``rare event'' filter, it is clear there are coherent structures on top of a background domain, but the four different structures can not be explicitly distinguished and identified. Thus a more detailed coherent structure filter using our unsupervised \localstate segmentation analysis is given in (c). Here states participating in the domain spacetime symmetry tiling are colored green, and other non-domain states which satisfy our definition for a coherent structure are colored according to the structure(s) they belong to. Interaction states not associated with domain or a coherent structure are in black. An outline of analogous results for vortex shedding is shown in (d)-(f). (d) A vortex street in the cloud layer over the arctic (Source: https://photojournal.jpl.nasa.gov/catalog/PIA03448). (e) The local statistical complexity of the vorticity field for a canonical vortex street simulation, taken from \cite{Jani07a}, analogous to the qualitative structure filter of (b). (f) Closer to the more detailed and principled coherent structure filter of (c) are the colored vortices displayed on the cover of the NERSC Cori HPC system. We emphasize the analogy is not that learning about coherent structures in CAs will give insight into fluid and climate structures. Rather, it is to illustrate how we foresee our approach will discover coherent structures in fluids and climate, in much the same way we can currently discover structures in CAs.}
   \label{fig}
\end{figure*}

\section{Towards Climate}
With consistent and readily interpretable results on cellular automata we are now working on generalizing to real-valued spatiotemporal systems, with specific emphasis on canonical fluid flows. Others have done preliminary work on this generalization, where an extra discretization (typically via clustering) step is needed during reconstruction \cite{Jani07a,Goer13a}. 

These groups have also used \localstates for coherent structure detection, including real-valued applications like fluids and even climate \cite{Jani07a}. However, they have all relied on the ``local statistical complexity'' \cite{Shal06a}, which is the point-wise entropy over \localstates. At best this is simply a qualitative filtering tool which aides in visual recognition of structures and at worst can give both false positive and false negative misidentification. We are the first to give a principled and rigorous method for coherent structure discovery and description using the \localstates, and are working to generalize this more detailed analysis to real-valued systems. In doing so we hope to move beyond the scope of data visualization these prior groups were working in, and facilitate novel scientific discovery, particularly in climate science. 

On the theory side, we must confirm our methods on known fluid structures. As the theory is founded in basic dynamical principles it is likely to apply without much modification in fluid systems. We will also begin to explore whether our methods can facilitate additional mechanistic insight beyond structure discovery. For example, whether there are any links between the \localstate analysis and thermodynamic considerations. 

On the implementation side, the computational costs of \localstate reconstruction in more complex systems will require fully-distributed execution on large HPC machines. This will certainly be the case for TB scale climate data sets we ultimately are interested in. As our primary objective is automated coherent structure discovery, moving from canonical fluid flows to large-scale climate data will largely be a matter of computational scaling. With access to HPC experts from the Intel Big Data Center and the NERSC Cori system at Lawrence Berkeley National Laboratory we feel well-positioned to tackle these computational challenges.

\section*{Acknowledgments}
Adam Rupe and Jim Crutchfield would like to acknowledge Intel for supporting the IPCC at UC Davis. Prabhat and Karthik Kashinath were supported by the Intel Big Data Center. This research is based upon work supported by, or in part by, the U. S. Army Research Laboratory and the U. S. Army Research Office under contract W911NF- 13-1-0390, and used resources of the National Energy Research Scientific Computing Center, a DOE Office of Science User Facility supported by the Office of Science of the U.S. Department of Energy under Contract No. DE-AC02-05CH11231. 

\bibliographystyle{ieeetr}
\bibliography{chaos,spacetime}

\end{document}